\documentclass[a4paper,12pt]{article}
\usepackage{amssymb}
\usepackage{amsmath}
\usepackage{epsfig}
\usepackage{subfigure}
\usepackage{graphics}

\usepackage{latexsym}
\usepackage{rotating}
\usepackage{titlesec}

\title{$(2+1)$-Dimensional Local and  Nonlocal Reductions of the Negative AKNS System: Soliton Solutions}

\author{Metin G\"{u}rses \thanks{gurses@fen.bilkent.edu.tr}\\
{\small Department of Mathematics, Faculty of Science}\\
{\small Bilkent University, 06800 Ankara - Turkey}\\
Asl{\i} Pekcan \thanks{Email:aslipekcan@hacettepe.edu.tr} \\
{\small Department of Mathematics, Faculty of Science} \\
{\small Hacettepe University, 06800 Ankara - Turkey}
}

\date{\nonumber}
\begin{document}
\maketitle
\date{\nonumber}
\newtheorem{thm}{Theorem}[section]
\newtheorem{Le}{Lemma}[section]
\newtheorem{defi}{Definition}[section]
\newtheorem{ex}{Example}[section]
\newtheorem{pro}{Proposition}[section]
\baselineskip 17pt

\numberwithin{equation}{section}

\begin{abstract}
We first construct a $(2+1)$-dimensional negative AKNS hierarchy and then we give all possible local and (discrete) nonlocal reductions of these
equations. We find Hirota bilinear forms of the negative AKNS hierarchy and give one- and two-soliton solutions. By using the soliton solutions of the negative AKNS hierarchy we find one-soliton solutions of the local and nonlocal reduced equations.\\

\end{abstract}

\section{Introduction}

Let ${\cal R}$ be the recursion operator of an integrable equation. Then the integrable hierarchy of equations are defined as
\begin{equation}\label{hier1}
v_{t_{n}}={\cal R}^n\, v_{x}\quad n=0,1,2,\ldots .
\end{equation}
In \cite{MGAP}, we proposed a  system of equations
\begin{equation}\label{compactform}
{\bf \cal R}[v_{t_{n}}-a{\cal R}^n \sigma_{0}]=b\sigma_1,\quad n=0,1,2,\ldots ,
\end{equation}
where  $\sigma_{0},\sigma_1$ are some classical symmetries of the same integrable equation. This hierarchy represents the negative hierarchy of the integrable system defined in (\ref{hier1}).
For some specific choices of the constants $a, b,$ and $\sigma_{0},\sigma_1$ we have studied the existence of  three-soliton solutions and Painlev{\'e} property of the KdV equation where the recursion operator is ${\cal R}=D^2+8v+4v_xD^{-1}$. The equation (\ref{compactform}) becomes the KdV(6) equation when $a=-1, b=0, n=1$ and by letting $v=u_x$ to get rid of nonlocal terms containing $D^{-1}$. We have also obtained $(2+1)$-extension of this equation, the $(2+1)$-KdV(6) equation  by choosing $a=-1,b=-1,n=1$, and $\sigma_{0}=v_{x}$, $\sigma_1=v_y$. The expanded form with $v=u_x$ of is the $(2+1)$-KdV(6) equation \cite{kar}
\begin{equation}\label{eqn}
u_{xxxt}+u_{xxxxxx}+40u_{xx}u_{xxx}+20u_xu_{xxxx}+8u_xu_{xt}+120u_x^2u_{xx}+4u_tu_{xx}+u_{xy}=0.
\end{equation}
We showed that $(2+1)$-KdV(6) equation possesses three-soliton solution having the same structure with the KdV equation's three-soliton solution and also Painlev{\'
e} property.

\noindent  By using our approach (\ref{compactform}), we obtain negative hierarchy of integrable equations which are nonlocal in general. Here nonlocality is due to the existence of the terms containing the operator $D^{-1}$. In the KdV case the nonlocal terms disappear by redefinition of the dynamical variable. This may not be possible for other integrable systems.

 A new type of nonlocal reductions are obtained by relating one of the dynamical variable to the time and space reflections of the other one which was first introduced by Ablowitz and Musslimani \cite{AbMu1}-\cite{AbMu3}. Ablowitz-Muslimani type of nonlocal reductions attracted many researchers \cite{fok}-\cite{GurPek2} to investigate new nonlocal integrable equations and find their solitonic solutions. These nonlocal integrable equations have been obtained by the nonlocal reductions of the AKNS and other systems of equations.  First example was the nonlocal nonlinear Schr\"{o}dinger (NLS) equation and then nonlocal modified KdV (mKdV) equation. Ablowitz and Musslimani proposed later some other nonlocal  integrable equations such as reverse space-time and reverse time nonlocal NLS equation, sine-Gordon equation, $(1+1)$- and $(2+1)$- dimensional three-wave interaction, Davey-Stewartson equation, derivative NLS equation, ST-symmetric nonlocal complex mKdV and mKdV equations arising from symmetry reductions of general AKNS scattering problem \cite{AbMu1}-\cite{AbMu3}. They discussed Lax pairs, an infinite number of conservation laws, inverse scattering transforms and found one-soliton solutions of these equations. Ma, Shen, and Zhu showed that ST-symmetric nonlocal complex mKdV equation is gauge equivalent to a spin-like model in Ref. \cite{ma}.  Ji and Zhu obtained soliton, kink, anti-kink, complexiton, breather, rogue-wave solutions, and nonlocalized solutions with singularities of ST-symmetric nonlocal mKdV equation through Darboux transformation and inverse scattering transform \cite{JZ1}, \cite{JZ2}. In \cite{Yang}, the authors showed that many nonlocal integrable equations like Davey-Stewartson equation, T-symmetric NLS equation, nonlocal derivative NLS equation, and ST-symmetric
complex mKdV equation can be converted to local integrable equations by simple variable transformations.

Recently we studied all possible nonlocal reductions of the AKNS system. We have obtained one-, two-, and three-soliton solutions of the nonlocal NLS \cite{GurPek1} and mKdV equations \cite{GurPek2}. We also studied nonlocal reductions of Fordy-Kulish \cite{GursesFK} and super integrable systems \cite{gur3}, \cite{GurPek3}.

In this work, by the use of the formula (\ref{compactform}) we obtain negative AKNS hierarchy denoted by AKNS($-n$) for $n=0,1,2, \ldots$ with one time $t$ and two space variables $x$ and $y$. This new system constitutes one of the few examples of $(2+1)$-dimensional integrable system of equations \cite{fok}, \cite{yulei}.  All these systems are nonlocal due to the term $D^{-1}$ in the recursion operator. We obtain the Hirota bilinear form of these systems and obtain one- and two-soliton solutions for $n=0,1,2$. We then find all possible local and nonlocal reductions of the  negative AKNS hierarchy for $n=0,1,2$. There are in total 30 reduced equations for $n=0,1,2$. All these equations constitute new examples of $(2+1)$-dimensional integrable system of equations \cite{fok}, \cite{yulei}. There exists only one type of local reductions where the second dynamical variable is related to the complex conjugation of the other variable. By the use of constraint equations we obtain one-soliton solutions of the local and nonlocal reduced equations from the one-soliton solutions of the negative AKNS system of equations. There are solutions which develop
singularities in a finite time and there are also solutions which are finite and bounded depending on the parameters of the one-soliton solutions.
\bigskip

\section{Negative AKNS System}

\noindent The AKNS hierarchy  \cite{AKNS} can be written as

$$u_{t_n}=\mathcal{R}^{n} u_x, \, (n=0,1,2,\ldots),\, u= \left( \begin{array}{c}
p  \\
q
 \end{array} \right) \, \mathrm{i.e.} \,  \left( \begin{array}{c}
p_{t_N}  \\
q_{t_N}

 \end{array} \right)= \mathcal{R}^{N-1} \left( \begin{array}{c}
p_x  \\
q_x
 \end{array} \right),
$$
\noindent where $\mathcal{R}$ is the recursion operator,

\[ \mathcal{R}=\left( \begin{array}{cc}
-pD^{-1}q+\frac{1}{2}D & -pD^{-1}p  \\
qD^{-1}q & qD^{-1}p-\frac{1}{2}D
 \end{array} \right).\]

\noindent Here $D$ is the total $x$-derivative and $D^{-1}=\int^x$ (standard anti-derivative).

\noindent Writing (\ref{compactform}) in the following form
\begin{equation}\label{AKNSextension}
\mathcal{R}(u_{t_n})-a\mathcal{R}^{n} (u_{x})=b\,u_y \ \mathrm{for} \ n=0,1,\ldots,
\end{equation}
where $u=\left( \begin{array}{c}
p  \\
q
 \end{array} \right),$  here $a,b$ are any constants, we obtain $(2+1)$-dimensional negative AKNS($-n$) systems for $n=0,1,2$.

 \vspace{0.5cm}
\noindent
{\bf (1) $(n=0)$ $(2+1)$-AKNS(0) System:}\\

\noindent
When $n=0$, Eq. (\ref{AKNSextension}) reduces to $\mathcal{R}(u_t)-au_x=bu_y$.
 This yields the system
 \begin{eqnarray}
 &&b p_{y}=\frac{1}{2}\,p_{tx}-a\, p_{x}-p D^{-1}\, (pq)_{t},\label{n=0p_y} \\
 &&b q_{y}=-\frac{1}{2}\,q_{tx}-a\, q_{x}+q D^{-1}\, (pq)_{t}.\label{n=0q_y}
 \end{eqnarray}

\vspace{0.5cm}
 \noindent
{\bf (2) $(n=1)$ $(2+1)$-AKNS(-1) System:}\\

\noindent When $n=1$, Eq. (\ref{AKNSextension}) reduces to $\mathcal{R}(u_t-au_x)=bu_y$. Letting $u_t-au_x=\omega$, where $\omega=\left( \begin{array}{c}
\omega_1  \\
\omega_2
 \end{array} \right)$ we have
  \begin{equation*}
 u_t-au_x=\omega, \quad  \quad  \mathcal{R}\omega=bu_y.
 \end{equation*}
 \noindent This yields the system
 \begin{eqnarray}
 \displaystyle \omega_1&=&p_t-ap_x \nonumber\\
 \omega_2&=&q_t-aq_x\nonumber\\
 bp_y&=&\frac{1}{2}\omega_{1,x}-pD^{-1}(q\omega_1+p\omega_2)\nonumber\\
 bq_y&=&-\frac{1}{2}\omega_{2,x}+qD^{-1}(q\omega_1+p\omega_2).
  \end{eqnarray}

\noindent Inserting $\omega_{1}$ and $\omega_{2}$ we obtain the system
 \begin{eqnarray}
 &&b p_{y}=\frac{1}{2}\,p_{tx}-\frac{a}{2}\, p_{xx}+a p^2\, q-p D^{-1}\, (pq)_{t},\label{n=1p_y} \\
 &&b q_{y}=-\frac{1}{2}\,q_{tx}+\frac{a}{2}\, q_{xx}-a p\, q^2+q D^{-1}\, (pq)_{t}.\label{n=1q_y}
 \end{eqnarray}

 \vspace{0.5cm}
\noindent
{\bf (3) $(n=2)$ $(2+1)$-AKNS(-2) System:}\\

\noindent When $n=2$, Eq. (\ref{AKNSextension}) reduces to $\mathcal{R}(u_t-a\mathcal{R}u_x)=bu_y$. Letting $u_t-a\mathcal{R}u_x=\omega$, where $\omega=\left( \begin{array}{c}
\omega_1  \\
\omega_2
 \end{array} \right)$ we have
  \begin{equation*}
 u_t-a\mathcal{R}u_x=\omega, \quad  \quad  \mathcal{R}\omega=bu_y.
 \end{equation*}
\noindent This yields the system
 \begin{eqnarray}\label{n=2}
 \displaystyle \omega_1&=&p_t-a(-p^2q+\frac{1}{2}p_{xx}) \nonumber\\
 \omega_2&=&q_t-a(pq^2-\frac{1}{2}q_{xx})\nonumber\\
 bp_y&=&\frac{1}{2}\omega_{1,x}-pD^{-1}(q\omega_1+p\omega_2)\nonumber\\
 bq_y&=&-\frac{1}{2}\omega_{2,x}+qD^{-1}(q\omega_1+p\omega_2).
  \end{eqnarray}
\noindent Inserting $\omega_{1}$ and $\omega_{2}$ we obtain the system
 \begin{eqnarray}
 &&b p_{y}=\frac{1}{2}\,p_{tx}-\frac{a}{4}\, p_{xxx}+\frac{3a}{2} p\, q\, p_{x}-p D^{-1}\, (pq)_{t},\label{n=2p_y} \\
 &&b q_{y}=-\frac{1}{2}\,q_{tx}-\frac{a}{4}\, q_{xxx}+\frac{3a}{2} p\, q\,q_{x}+q D^{-1}\, (pq)_{t}.\label{n=2q_y}
 \end{eqnarray}

 \section{Hirota Method for Negative AKNS System}

To obtain the Hirota bilinear form for the negative AKNS($-n$) system, with $n=0, 1,$ and $n=2$, we let

\begin{equation}
p=\frac{g}{f},\quad  q=\frac{h}{f}, \label{hir1}
\end{equation}
and

 \begin{equation}
 \frac{gh}{f^2}=- \left(\frac{f_{x}}{f} \right)_{x}. \label{hir2}
 \end{equation}

  \vspace{0.5cm}
 \noindent
 {\bf (1)  $(n=0)$ Hirota Bilinear Form for $(2+1)$-AKNS(0) System:}

\noindent Using (\ref{hir1}) and (\ref{hir2}) in Eqs. (\ref{n=0p_y}) and (\ref{n=0q_y}) give
 \begin{eqnarray}
 &&b(f g_{y}-g f_{y})=\frac{1}{2}\, (f g_{tx}-g_{t} f_{x}-g_{x} f_{t}+g f_{tx})-a (f g_{x}-g f_{x} ), \\
 &&b(f h_{y}-h f_{y})=-\frac{1}{2}\, (f h_{tx}-h_{t} f_{x}-h_{x} f_{t}+h f_{tx} )-a(f h_{x}-h f_{x} ).
 \end{eqnarray}
 Hence we obtain the Hirota bilinear form as
 \begin{eqnarray}
&& P_1(D)\{ g \cdot f\}\equiv (b D_{y}-\frac{1}{2}\, D_{t} D_{x}+a\,D_{x} )\{ g \cdot f\}=0, \label{n=0hirota1}\\
&& P_2(D)\{ h \cdot f\}\equiv (b D_{y}+\frac{1}{2}\, D_{t} D_{x}+a\, D_{x}) \{ h \cdot f\}=0,\label{n=0hirota2}\\
&& P_3(D)\{ f \cdot f\}\equiv D_{x}^2 \{f \cdot f\}=-2 g h. \label{n=0hirota3}
\end{eqnarray}

 \vspace{0.5cm}
 \noindent
 {\bf (2) $(n=1)$ Hirota Bilinear Form for $(2+1)$-AKNS(-1) System:}

\noindent Using (\ref{hir1}) and (\ref{hir2}) in Eqs. (\ref{n=1p_y}) and (\ref{n=1q_y}) give
 \begin{eqnarray}
 &&b(f g_{y}-g f_{y})=\frac{1}{2}\, (f g_{tx}-g_{t} f_{x}-g_{x} f_{t}+g f_{tx})-\frac{a}{2} (f g_{xx}-2 f_{x} g_{x}+g f_{xx} ), \\
 &&b(f h_{y}-h f_{y})=-\frac{1}{2}\, (f h_{tx}-h_{t} f_{x}-h_{x} f_{t}+h f_{tx} )+\frac{a}{2} (f h_{xx}-2 f_{x} h_{x}+h f_{xx} ).
 \end{eqnarray}
 Hence we obtain the Hirota bilinear form as
 \begin{eqnarray}
&& P_1(D)\{ g \cdot f\}\equiv (b D_{y}-\frac{1}{2}\, D_{t} D_{x}+\frac{a}{2} D_{x}^2 )\{ g \cdot f\}=0, \label{n=1hirota1}\\
&& P_2(D)\{ h \cdot f\}\equiv (b D_{y}+\frac{1}{2}\, D_{t} D_{x}-\frac{a}{2} D_{x}^2 ) \{ h \cdot f\}=0,\label{n=1hirota2}\\
&& P_3(D)\{ f \cdot f\}\equiv D_{x}^2 \{f \cdot f\}=-2 g h. \label{n=1hirota3}
\end{eqnarray}

\vspace{0.5cm}
 \noindent
 {\bf (3) $(n=2)$ Hirota Bilinear Form for $(2+1)$-AKNS(-2) System:}

\noindent Using (\ref{hir1}) and (\ref{hir2}) in Eqs. (\ref{n=2p_y}) and (\ref{n=2q_y}) yield
 \begin{align}
&4b(f g_{y}-g f_{y})= 2 (f g_{tx}-g_{t} f_{x}-g_{x} f_{t}+g f_{tx} )-a(f g_{xxx}+3 f_{xx} g_{x}-3 g_{xx} f_{x}-g f_{xxx}), \\
 &4b(f h_{y}-h f_{y})=-2 (f h_{tx}-h_{t} f_{x}-h_{x} f_{t}+h f_{tx})
 -a(f h_{xxx}-3 f_{x} h_{xx}+3 h_{x} f_{xx}-h f_{xxx} ).
 \end{align}
 Hence we obtain the Hirota bilinear form as
 \begin{eqnarray}
&& P_1(D) \{ g \cdot f \}\equiv (b D_{y}-\frac{1}{2} D_{t} D_{x}+\frac{a}{4} D^3_{x} ) \{g \cdot f\}=0,\label{n=2hirota1} \\
&&  P_2(D) \{ g \cdot f \}\equiv (b D_{y}+\frac{1}{2} D_{t} D_{x}+\frac{a}{4} D^3_{x} ) \{h \cdot f\}=0,\label{n=2hirota2}\\
&&  P_3(D) \{ f \cdot f \}\equiv D_{x}^2 \{f \cdot f \}=-2 g h.\label{n=2hirota3}
\end{eqnarray}

\noindent After having Hirota bilinear forms (\ref{n=0hirota1})-(\ref{n=0hirota3}), (\ref{n=1hirota1})-(\ref{n=1hirota3}), and (\ref{n=2hirota1})-(\ref{n=2hirota3}), next step is to find the functions $g$, $h$, and $f$ by using the Hirota method (see Sec. VI).

\section{Local Reductions}

It is straightforward to show that there exist no consistent local reductions in the form of  $q(x,y,t)=\sigma\, p(x,y,t)$  for all $n=0, 1, 2$. Here we will give  the local reductions in the form of $q(x,y,t)=\sigma\, \bar{p}(x,y,t)$ for all $n=0, 1, 2$ where $\sigma$ is any real constant.


\vspace{0.5cm}
\noindent
{\bf (1) Local Reductions for the System  $n=0$:}\\

Let $q(x,y,t)=\sigma\, \bar{p}(x,y,t)$ then two coupled equations (\ref{n=0p_y}) and (\ref{n=0q_y}) reduce consistently to the following single equation
\begin{equation}
b p_{y}=\frac{1}{2}\,p_{tx}-a\, p_{x}-\sigma\,p D^{-1}\, (p\, \bar{p})_{t}\label{localreducedn=0},
\end{equation}
where $\sigma$ is any real constant and a bar over a letter denotes complex conjugation. Here $a$ and $b$ are pure imaginary numbers.

\vspace{0.5cm}
\noindent
{\bf (2) Local Reductions for the System  $n=1$:}\\

Let $q(x,y,t)=\sigma\, \bar{p}(x,y,t)$ then two coupled equations (\ref{n=1p_y}) and (\ref{n=1q_y}) reduce consistently to the following single equation
\begin{equation}
b p_{y}=\frac{1}{2}\,p_{tx}-\frac{a}{2}\, p_{xx}+a \sigma\,p^2\, \bar{p}-\sigma\, p D^{-1}\, (p\, \bar{p})_{t},\label{localreducedn=1}
\end{equation}
where $\sigma$ is any real constant and a bar over a letter denotes complex conjugation. Here $a$ is a real and $b$ is a pure imaginary number.

\vspace{0.5cm}
\noindent
{\bf (3) Local Reductions for the System  $n=2$:}\\

 Let $q(x,y,t)=\sigma\, \bar{p}(x,y,t)$ then two coupled equations (\ref{n=2p_y}) and (\ref{n=2q_y}) reduce consistently to the following single equation
\begin{equation}
b p_{y}=\frac{1}{2}\,p_{tx}-\frac{a}{4}\, p_{xxx}+\frac{3a}{2} \sigma\, p\, \bar{p}\, p_{x}-\sigma\,p D^{-1}\, (p \bar{p})_{t},\label{localreducedn=2}
\end{equation}
where $\sigma$ is any real constant and a bar over a letter denotes complex conjugation. Here $a$ and $b$ are pure imaginary numbers.

\vspace{0.3cm}

\section{Nonlocal Reductions}

In order to have consistent nonlocal reductions we use the following representation for $D^{-1}$
\begin{equation}
D^{-1}\,F=\frac{1}{2}\, \left(\int_{-\infty}^{x}-\int_{x}^{\infty}\,\right) F(x^{\prime},y,t) dx^{\prime}.
\end{equation}
We define the quantity $\rho(x,y,t)$ which is invariant under the discrete transformations $x \to \epsilon_{1} x$, $y \to \epsilon_{2} y,$ and $t \to \epsilon_{3} t$ as
\begin{equation}
\rho (x,y,t)=D^{-1}\,p\, p^{\epsilon} \equiv \left(\int_{-\infty}^{x}-\int_{x}^{\infty}\,\right) p(x^{\prime},y,t)\, p(\epsilon_{1}\, x^{\prime}, \epsilon_{2}\, y, \epsilon_{3}\, t)\, dx^{\prime},
\end{equation}
where $\epsilon_{1}^2=\epsilon_{2}^2=\epsilon_{3}^2=1$. It is easy to show that
\begin{equation}
\rho(\epsilon_{1}\, x, \epsilon_{2}\, y, \epsilon_{3}\, t)=\epsilon_{1}\, \rho(x,y,t).
\end{equation}

\vspace{0.5cm}
\noindent
{\bf (1) Nonlocal Reductions for the System $n=0$:}\\

\noindent
{\bf (a)} Let $q(x,y,t)=\sigma\, p(\epsilon_{1} x,\epsilon_{2} y, \epsilon_{3} t)$  then two coupled equations (\ref{n=0p_y}) and (\ref{n=0q_y}) reduce consistently to the following single equation
\begin{equation}
b p_{y}=\frac{1}{2}\,p_{tx}-a\, p_{x}-\sigma\,p D^{-1}\, (p\, p^{\epsilon})_{t},\label{nonlocalredn=0a}
\end{equation}
where $\sigma$ is any real constant and $p^{\epsilon}= p(\epsilon_{1} x,\epsilon_{2} y, \epsilon_{3} t)$. The above reduced equation is valid only when $\epsilon_{3}=-1$ and $\epsilon_{1}\, \epsilon_{2}=1$. We have only two possible cases: $p^{\epsilon}=p(x,y,-t)$ and $p^{\epsilon}=p(-x,-y,-t)$ for time reversal and time and space reversals respectively.

\vspace{0.3cm}
\noindent
{\bf (b)} Let $q(x,y,t)=\sigma\, \bar{p}(\epsilon_{1} x,\epsilon_{2} y, \epsilon_{3} t)$  then two coupled equations (\ref{n=0p_y}) and (\ref{n=0q_y}) reduce consistently to the following single equation
\begin{equation}
b p_{y}=\frac{1}{2}\,p_{tx}-a\, p_{x}-\sigma\,p D^{-1}\, (p\, \bar{p}^{\epsilon})_{t},\label{nonlocalredn=0b}
\end{equation}
where $\sigma$ is any real constant. This reduction is valid only when
\begin{equation}\label{addcondn=0}
\epsilon_{1}\, \epsilon_{2}\, \epsilon_{3}\, \bar{b}=-b,~~~~ \epsilon_{3}\, \bar{a}=-a.
\end{equation}
In this case we have seven different time and space reversals:

\vspace{0.2cm}
\noindent
(i) $p^{\epsilon}(x,y,t)=p(-x,y,t)$, where $a$ is pure imaginary and $b$ is real.\\
(ii) $p^{\epsilon}(x,y,t)=p(x,-y,t)$, where $a$ is pure imaginary and $b$ is real.\\
(iii. $p^{\epsilon}(x,y,t)=p(x,y,-t)$, where $a$ and $b$ are real.\\
(iv) $p^{\epsilon}(x,y,t)=p(-x,-y,t)$, where $a$ and $b$ are pure imaginary. \\
(v) $p^{\epsilon}(x,y,t)=p(-x,y,-t)$, where $a$ is real and $b$ is pure imaginary.\\
(vi) $p^{\epsilon}(x,y,t)=p(x,-y,-t)$, where $a$ is real and  $b$ is pure imaginary. \\
(vii) $p^{\epsilon}(x,y,t)=p(-x,-y,-t)$, where $a$ and $b$ are real.\\

\noindent
Each case above gives a nonlocal equation in the form of (\ref{nonlocalredn=0b}) in 2+1 dimensions.

\vspace{0.5cm}
\noindent
{\bf (2) Nonlocal Reductions for the System $n=1$:}\\

\noindent
{\bf (a)} Let $q(x,y,t)=\sigma\, p(\epsilon_{1} x,\epsilon_{2} y, \epsilon_{3} t)$  then two coupled equations (\ref{n=1p_y}) and (\ref{n=1q_y}) reduce consistently to the following single equation
\begin{equation}
b p_{y}=\frac{1}{2}p_{tx}-\frac{a}{2} p_{xx}+a\sigma p^2 p^{\epsilon}-\sigma p D^{-1} (p p^{\epsilon})_{t},\label{nonlocalredn=1a}
\end{equation}
where $\sigma$ is any real constant and $p^{\epsilon}= p(\epsilon_{1} x,\epsilon_{2} y, \epsilon_{3} t)$. The above reduced equation is valid only when $\epsilon_{2}=-1$ and $\epsilon_{1} \epsilon_{3}=1$. We have only two possible cases: $p^{\epsilon}=p(x,-y,t)$ and $p^{\epsilon}=p(-x,-y,-t)$ for space reversal and time and space reversals respectively.

\vspace{0.3cm}
\noindent
{\bf (b)} Let $q(x,y,t)=\sigma\, \bar{p}(\epsilon_{1} x,\epsilon_{2} y, \epsilon_{3} t)$  then two coupled equations (\ref{n=1p_y}) and (\ref{n=1q_y}) reduce consistently to the following single equation
\begin{equation}
b p_{y}=\frac{1}{2}p_{tx}-\frac{a}{2} p_{xx}+a\sigma p^2 \bar{p}^{\epsilon}-\sigma p D^{-1} (p \bar{p}^{\epsilon})_{t}, \label{nonlocalredn=1b}
\end{equation}
where $\sigma$ is any real constant. This reduction is valid only when
\begin{equation}\label{addcondn=1}
\epsilon_{1}\, \epsilon_{2}\, \epsilon_{3}\, \bar{b}=-b,~~~~\epsilon_{1}\, \epsilon_{3}\, \bar{a}=a.
\end{equation}
In this case we have seven different time and space reversals:

\vspace{0.2cm}
\noindent
(i) $p^{\epsilon}(x,y,t)=p(-x,y,t)$, where $a$ is pure imaginary and $b$ is real.\\
(ii) $p^{\epsilon}(x,y,t)=p(x,-y,t)$, where $a$ and $b$ are real.\\
(iii) $p^{\epsilon}(x,y,t)=p(x,y,-t)$, where $a$ is pure imaginary and $b$ are real.\\
(iv) $p^{\epsilon}(x,y,t)=p(-x,-y,t)$, where $a$ and $b$ are pure imaginary. \\
(v) $p^{\epsilon}(x,y,t)=p(-x,y,-t)$, where $a$ is real and $b$ is pure imaginary.\\
(vi) $p^{\epsilon}(x,y,t)=p(x,-y,-t)$, where $a$ and  $b$ are pure imaginary. \\
(vii) $p^{\epsilon}(x,y,t)=p(-x,-y,-t)$, where $a$ and $b$ are real.\\

\noindent
Each case above gives a nonlocal equation in the form of (\ref{nonlocalredn=1b}) in 2+1 dimensions.

\vspace{0.5cm}
\noindent
{\bf (3) Nonlocal Reductions for the System $n=2$:}\\

\noindent
{\bf (a)} Let $q(x,y,t)=\sigma\, p(\epsilon_{1} x,\epsilon_{2} y, \epsilon_{3} t)$  then two coupled equations (\ref{n=2p_y}) and (\ref{n=2q_y}) reduce consistently to the following single equation
\begin{equation}
b p_{y}=\frac{1}{2}p_{tx}-\frac{a}{4} p_{xxx}+\frac{3a}{2}\sigma p p^{\epsilon}p_x-\sigma p D^{-1} (p p^{\epsilon})_{t},\label{nonlocalredn=2a}
\end{equation}
where $\sigma$ is any real constant and $p^{\epsilon}= p(\epsilon_{1} x,\epsilon_{2} y, \epsilon_{3} t)$. The above reduced equation is valid only when $\epsilon_{3}=-1$ and $\epsilon_{1} \epsilon_{2}=1$. We have only two possible cases: $p^{\epsilon}=p(x,y,-t)$ and $p^{\epsilon}=p(-x,-y,-t)$ for time reversal and time and space reversals respectively.

\vspace{0.3cm}
\noindent
{\bf (b)} Let $q(x,y,t)=\sigma\, \bar{p}(\epsilon_{1} x,\epsilon_{2} y, \epsilon_{3} t)$  then two coupled equations (\ref{n=1p_y}) and (\ref{n=1q_y}) reduce consistently to the following single equation
\begin{equation}
b p_{y}=\frac{1}{2}p_{tx}-\frac{a}{4} p_{xxx}+\frac{3a}{2}\sigma p \bar{p}^{\epsilon}p_x-\sigma p D^{-1} (p \bar{p}^{\epsilon})_{t},\label{nonlocalredn=2b}
\end{equation}
where $\sigma$ is any real constant. This reduction is valid only when
\begin{equation}\label{addcondn=2}
\epsilon_{1}\, \epsilon_{2}\, \epsilon_{3}\, \bar{b}=-b,~~~~\epsilon_{3}\, \bar{a}=-a.
\end{equation}
In this case we have seven different time and space reversals:

\vspace{0.2cm}
\noindent
(i) $p^{\epsilon}(x,y,t)=p(-x,y,t)$, where $a$ is pure imaginary and $b$ is real.\\
(ii) $p^{\epsilon}(x,y,t)=p(x,-y,t)$, where $a$ is pure imaginary and $b$ is real.\\
(iii) $p^{\epsilon}(x,y,t)=p(x,y,-t)$, where $a$ and $b$ are real.\\
(iv) $p^{\epsilon}(x,y,t)=p(-x,-y,t)$, where $a$ and $b$ are pure imaginary. \\
(v) $p^{\epsilon}(x,y,t)=p(-x,y,-t)$, where $a$ is real and $b$ is pure imaginary.\\
(vi) $p^{\epsilon}(x,y,t)=p(x,-y,-t)$, where $a$ is real and  $b$ is pure imaginary. \\
(vii) $p^{\epsilon}(x,y,t)=p(-x,-y,-t)$, where $a$ and $b$ are real.\\

\noindent
Each case above gives a nonlocal equation in the form of (\ref{nonlocalredn=2b}) in 2+1 dimensions. At the end we obtain 27 nonlocal equations from negative AKNS hierarchy in 2+1 dimensions.

\vspace{0.2cm}
\noindent
\textbf{Remark.} In all the above nonlocal equations we can use $D^{-1}=\int^x$ when there exist only $y$ and $t$ reversals, $p^{\epsilon}=p(x, \epsilon_{2}\,y, \epsilon_{3}\, t)$.

\section{Soliton Solutions for Negative AKNS Hierarchy}

\subsection{One-Soliton Solution of $(2+1)$-AKNS($-n$) System $(n=0, 1, 2)$}

Here we will present how to find one-soliton solution of $(2+1)$-AKNS(0) system. For $n=1$ and $n=2$ the steps for finding one-soliton solution are same with $n=0$ case except the dispersion relations.

\noindent Consider the system (\ref{n=0hirota1})-(\ref{n=0hirota3}). To find one-soliton solution we use the following expansions for the functions $g$, $h$, and $f$,
\begin{equation}\label{n=0expansion}
g=\varepsilon g_1, \quad h=\varepsilon h_1, \quad f=1+\varepsilon^2 f_2,
\end{equation}
where
\begin{equation}\label{g_1h_1n=0}
g_1=e^{\theta_1}, \quad h_1=e^{\theta_2}, \quad \theta_i=k_ix+\tau_i y+\omega_i t+\delta_i,\, i= 1, 2.
\end{equation}
When we substitute (\ref{n=0expansion}) into the equations (\ref{n=0hirota1})-(\ref{n=0hirota3}), we obtain the
coefficients of $\varepsilon$ as
\begin{eqnarray}\displaystyle
&&P_1(D)\{g_1\cdot 1\}=bg_{1,y}-\frac{1}{2}g_{1,xt}+ag_{1,x}=0,\\
&&P_2(D)\{h_1\cdot 1\}=bh_{1,y}+\frac{1}{2}h_{1,xt}+ah_{1,x}=0,
\end{eqnarray}
yielding the dispersion relations
\begin{equation}\label{dispersionn=0ONE}\displaystyle
\tau_1=\frac{1}{b}(\frac{1}{2}k_1\omega_1-ak_1),\quad \tau_2=\frac{1}{b}(-\frac{1}{2}k_2\omega_2-ak_2).
\end{equation}
From the coefficient of $\varepsilon^2$
\begin{equation}
f_{2,xx}=-g_1h_1,
\end{equation}
we obtain the function $f_2$ as
\begin{equation}\label{n=0f_2}
\displaystyle f_2=-\frac{ e^{(k_1+k_2)x+(\tau_1+\tau_2)y+(\omega_1+\omega_2)t+\delta_1+\delta_2} }{(k_1+k_2)^2}.
\end{equation}
The coefficients of $\varepsilon^3$ vanish with the dispersion relations and (\ref{n=0f_2}). From the coefficient of
$\varepsilon^4$ we get
\begin{equation}
f_2f_{2,xx}-f_{2,x}^2=0,
\end{equation}
and this equation also vanishes directly due to the dispersion relations and (\ref{n=0f_2}). Without loss of generality let us also take $\varepsilon=1$. Hence a pair of solutions of $(2+1)$-AKNS(0) system (\ref{n=0p_y})-(\ref{n=0q_y}) is given by $(p(x, y, t), q(x, y, t))$ where
\begin{equation}\label{n=0systemonesol}
\displaystyle p(x, y, t)=\frac{e^{\theta_1}}{1+Ae^{\theta_1+\theta_2}}, \quad \quad q(x, y, t)=\frac{e^{\theta_2}}{1+Ae^{\theta_1+\theta_2}},
\end{equation}
with $\theta_i=k_ix+\tau_iy+\omega_it+\delta_i$, $i=1,2$, $\tau_1=\frac{1}{b}(\frac{1}{2}k_1\omega_1-ak_1)$, $ \tau_2=\frac{1}{b}(-\frac{1}{2}k_2\omega_2-ak_2)$, and $ A=-\frac{1}{(k_1+k_2)^2}$. Here $k_{i}$, $\omega_{i}$, and $\delta_{i}$, $i=1, 2$  are arbitrary complex numbers.

\bigskip

\noindent For $n=1$ that is for the system (\ref{n=1p_y})-(\ref{n=1q_y}) one-soliton solution is given by (\ref{n=0systemonesol}) where $\theta_i=k_ix+\tau_iy+\omega_it+\delta_i$, $i=1,2$ with
\begin{equation}\label{dispersionn=1ONE}
\displaystyle \tau_1=\frac{1}{b}(\frac{1}{2}k_1\omega_1-\frac{a}{2}k_1^2),\quad
 \tau_2=\frac{1}{b}(-\frac{1}{2}k_2\omega_2+\frac{a}{2}k_2^2).
\end{equation}

\noindent For $n=2$ that is for the system (\ref{n=2p_y})-(\ref{n=2q_y}) one-soliton solution is again given by (\ref{n=0systemonesol}) where $\theta_i=k_ix+\tau_iy+\omega_it+\delta_i$, $i=1,2$ with
\begin{equation}\label{dispersionn=2ONE}
\displaystyle \tau_1=\frac{1}{b}(\frac{1}{2}k_1\omega_1-\frac{a}{4}k_1^3), \quad
 \tau_2=\frac{1}{b}(-\frac{1}{2}k_2\omega_2-\frac{a}{4}k_2^3).
\end{equation}

\subsection{Two-Soliton Solution of $(2+1)$-AKNS($-n$) System $(n=0, 1, 2)$}

 Here as in the previous section, we will only deal with $(2+1)$-AKNS(0) system and find two-soliton solution of this system.
 For $n=1$ and $n=2$ we have the same form of two-soliton solution only with difference of the dispersion relations.

\noindent Consider the system (\ref{n=0hirota1})-(\ref{n=0hirota3}). For two-soliton solution, we take
\begin{equation}\label{expansionn=02}
g=\varepsilon g_1+\varepsilon^3 g_3, \quad h=\varepsilon h_1+\varepsilon^3 h_3, \quad f=1+\varepsilon^2 f_2+\varepsilon^4 f_4,
\end{equation}
where
\begin{equation}\label{funcg_1h_1}
g_1=e^{\theta_1}+e^{\theta_2}, \quad h_1=e^{\eta_1}+e^{\eta_2},
\end{equation}
with $\theta_i=k_i x+\tau_i y+\omega_i t+\delta_i$, $\eta_i=\ell_ix+s_iy+m_it+\alpha_i$ for $i=1, 2$. When we insert above expansions into (\ref{n=0hirota1})-(\ref{n=0hirota3}),
we get the coefficients of $\varepsilon^n$, $1\leq n \leq 8$ as
\begin{align}
\varepsilon:&\, bg_{1,y}-\frac{1}{2}g_{1,xt}+ag_{1,x}=0,\label{twod1} \\
&\, bh_{1,y}+\frac{1}{2}h_{1,xt}+ah_{1,x}=0,\label{twod2} \\
\varepsilon^2:&\, f_{2,xx}+g_1h_1=0,\label{twod3}\\
\varepsilon^3:&\, b(g_{1,y}f_2-g_1f_{2,y})-\frac{1}{2}(g_{1,xt}f_2-g_{1,t}f_{2,x}-g_{1,x}f_{2,t}+g_1f_{2,xt})+a(g_{1,x}f_2-g_1f_{2,x})\nonumber\\
&\, +bg_{3,y}-\frac{1}{2}g_{3,xt}+ag_{3,x}=0,\label{twod4}\\
&\, b(h_{1,y}f_2-h_1f_{2,y})+\frac{1}{2}(h_{1,xt}f_2-h_{1,t}f_{2,x}-h_{1,x}f_{2,t}+h_1f_{2,xt})+a(h_{1,x}f_2-h_1f_{2,x})\nonumber\\
&\, +bh_{3,y}+\frac{1}{2}h_{3,xt}+ah_{3,x}=0,\label{twod5}\\
\varepsilon^4:&\, f_2f_{2,xx}-f_{2,x}^2+f_{4,xx}+g_1h_3+g_3h_1=0,\label{twod6}
\end{align}
\begin{align}
\varepsilon^5:&\,  b(g_{1,y}f_4-g_1f_{4,y})-\frac{1}{2}(g_{1,xt}f_4-g_{1,t}f_{4,x}-g_{1,x}f_{4,t}+g_1f_{4,xt})+a(g_{1,x}f_4-g_1f_{4,x})\nonumber\\
&\, + b(g_{3,y}f_2-g_3f_{2,y})-\frac{1}{2}(g_{3,xt}f_2-g_{3,t}f_{2,x}-g_{3,x}f_{2,t}+g_3f_{2,xt})+a(g_{3,x}f_2-g_3f_{2,x})=0,\label{twod7}\\
&\, b(h_{1,y}f_4-h_1f_{4,y})+\frac{1}{2}(h_{1,xt}f_4-h_{1,t}f_{4,x}-h_{1,x}f_{4,t}+h_1f_{4,xt})+a(h_{1,x}f_4-h_1f_{4,x})\nonumber\\
&\, + b(h_{3,y}f_2-h_3f_{2,y})+\frac{1}{2}(h_{3,xt}f_2-h_{3,t}f_{2,x}-h_{3,x}f_{2,t}+h_3f_{2,xt})+a(h_{3,x}f_2-h_3f_{2,x})=0,\label{twod8}\\
\varepsilon^6:&\, f_{2,xx}f_4-2f_{2,x}f_{4,x}+f_2f_{4,xx}+g_3h_3=0,\label{twod9}\\
\varepsilon^7:&\,  b(g_{3,y}f_4-g_3f_{4,y})-\frac{1}{2}(g_{3,xt}f_4-g_{3,t}f_{4,x}-g_{3,x}f_{4,t}+g_3f_{4,xt})+a(g_{3,x}f_4-g_3f_{4,x})=0,\label{twod10}\\
&\, b(h_{3,y}f_4-h_3f_{4,y})+\frac{1}{2}(h_{3,xt}f_4-h_{3,t}f_{4,x}-h_{3,x}f_{4,t}+h_3f_{4,xt})+a(h_{3,x}f_4-h_3f_{4,x})=0.\label{twod11}
\\
\varepsilon^8:&\, f_4f_{4,xx}-f_{4,x}^2=0.\label{twod12}
\end{align}

\noindent The equations (\ref{twod1}) and (\ref{twod2}) give the dispersion relations
\begin{equation}\label{dispersiontwo}\displaystyle
\tau_i=\frac{1}{b}(\frac{1}{2}k_i\omega_i-ak_i), \quad s_i=\frac{1}{b}(-\frac{1}{2}\ell_im_i-a\ell_i), \quad i=1, 2.
\end{equation}
From the coefficient of $\varepsilon^2$ we obtain the function $f_2$,
\begin{equation}\displaystyle
f_2= e^{\theta_1+\eta_1+\alpha_{11}}+e^{\theta_1+\eta_2+\alpha_{12}}+e^{\theta_2+\eta_1+\alpha_{21}}+e^{\theta_2+\eta_2+\alpha_{22}}=\sum_{1\leq i,j\leq 2}e^{\theta_i+\eta_j+\alpha_{ij}},
\end{equation}
where
\begin{equation}
\displaystyle e^{\alpha_{ij}}=-\frac{1}{(k_i+\ell_j)^2},\, 1\leq i,j\leq 2.
\end{equation}

\noindent The equations (\ref{twod4}) and (\ref{twod5}) give the functions $g_3$ and $h_3$,
\begin{equation}\label{g_3h_3}
g_3=A_1e^{\theta_1+\theta_2+\eta_1}+A_2e^{\theta_1+\theta_2+\eta_2},\quad h_3=B_1e^{\theta_1+\eta_1+\eta_2}+B_2e^{\theta_2+\eta_1+\eta_2},
\end{equation}
where
\begin{equation}\label{gamma_ibeta_i}
\displaystyle A_i=-\frac{(k_1-k_2)^2}{(k_1+\ell_i)^2(k_2+\ell_i)^2}, \quad B_i=-\frac{(\ell_1-\ell_2)^2}{(\ell_1+k_i)^2(\ell_2+k_i)^2},\, i=1, 2.
\end{equation}

\noindent The equation (\ref{twod6}) yields the function $f_4$ as
\begin{equation}
f_4=Me^{\theta_1+\theta_2+\eta_1+\eta_2},
\end{equation}
where
\begin{equation}\label{M}
\displaystyle M=\frac{(k_1-k_2)^2(l_1-l_2)^2}{(k_1+l_1)^2(k_1+l_2)^2(k_2+l_1)^2(k_2+l_2)^2}.
\end{equation}
Other equations (\ref{twod7})-(\ref{twod12}) vanish immediately by the dispersion relations (\ref{dispersiontwo}) and the functions
$f_2, f_4, F_3$, and $G_3$.

\noindent Let us also take $\varepsilon=1$. Then two-soliton solution of the system (\ref{n=0p_y})-(\ref{n=0q_y}) is given with the pair $(p(x, y, t), q(x, y, t))$,
\begin{align}
\displaystyle& p(x,y,t)=\frac{e^{\theta_1}+e^{\theta_2}+A_1e^{\theta_1+\theta_2+\eta_1}+A_2e^{\theta_1+\theta_2+\eta_2}}
{1+e^{\theta_1+\eta_1+\alpha_{11}}+e^{\theta_1+\eta_2+\alpha_{12}}+e^{\theta_2+\eta_1+\alpha_{21}}+e^{\theta_2+\eta_2+\alpha_{22}}
+Me^{\theta_1+\theta_2+\eta_1+\eta_2}},\label{n=0twosolp(x,y,t)}\\
&q(x,y,t)=\frac{e^{\eta_1}+e^{\eta_2}+B_1e^{\theta_1+\eta_1+\eta_2}+B_2e^{\theta_2+\eta_1+\eta_2}}
{1+e^{\theta_1+\eta_1+\alpha_{11}}+e^{\theta_1+\eta_2+\alpha_{12}}+e^{\theta_2+\eta_1+\alpha_{21}}+e^{\theta_2+\eta_2+\alpha_{22}}
+Me^{\theta_1+\theta_2+\eta_1+\eta_2}},\label{n=0twosolq(x,y,t)}
\end{align}
with $\displaystyle \theta_i=k_ix+\tau_iy+\omega_i t+\delta_i$, $\displaystyle \eta_i=\ell_ix+s_iy+m_it+\alpha_i$ for $i=1, 2$ with
the dispersion relations $\tau_i=\frac{1}{b}(\frac{1}{2}k_i\omega_i-ak_i)$, $s_i=\frac{1}{b}(-\frac{1}{2}\ell_im_i-a\ell_i)$, $i=1, 2$. Here $k_{i}$, $\ell_{i}, \omega_i, m_i, \delta_{i}$, and $\alpha_{i}$, $i=1, 2$ are arbitrary complex numbers.

\bigskip

\noindent For $n=1$ i.e. for the system (\ref{n=1p_y})-(\ref{n=1q_y}) two-soliton solution is given by (\ref{n=0twosolp(x,y,t)})-(\ref{n=0twosolq(x,y,t)}) where $ \theta_i=k_ix+\tau_iy+\omega_i t+\delta_i$, $ \eta_i=\ell_ix+s_iy+m_it+\alpha_i$ for $i=1, 2$ with
the dispersion relations
\begin{equation}\displaystyle
\tau_i=\frac{1}{b}(\frac{1}{2}k_i\omega_i-\frac{a}{2}k_i^2),\quad s_i=\frac{1}{b}(-\frac{1}{2}\ell_im_i+\frac{a}{2}\ell_i^2), i=1, 2.
\end{equation}

\noindent For $n=2$ that is for the system (\ref{n=2p_y})-(\ref{n=2q_y}) two-soliton solution is also given by (\ref{n=0twosolp(x,y,t)})-(\ref{n=0twosolq(x,y,t)}) where $ \theta_i=k_ix+\tau_iy+\omega_i t+\delta_i$, $\eta_i=\ell_ix+s_iy+m_it+\alpha_i$ for $i=1, 2$ with the dispersion relations
\begin{equation}\displaystyle
\tau_i=\frac{1}{b}(\frac{1}{2}k_i\omega_i-\frac{a}{4}k_i^3),\quad s_i=\frac{1}{b}(-\frac{1}{2}\ell_im_i-\frac{a}{4}\ell_i^3), i=1, 2.
\end{equation}

\section{Soliton Solutions of Reduced Equations}
In our studies of nonlocal NLS and nonlocal mKdV equations we introduced a general method \cite{GurPek1}-\cite{GurPek3} to obtain soliton solutions of nonlocal integrable equation. This method consists of three main steps:

\begin{itemize}
\item Find a consistent reduction formula which reduces the integrable system of equations to integrable nonlocal equations.
\item Find soliton solutions of the system of equations by use of the Hirota bilinear method or by inverse scattering transform technique, or by use of Darboux Transformation.
\item Use the reduction formulas on the soliton solutions of the system of equations to obtain the soliton solutions of the reduced nonlocal equations. By this way one obtains many different relations among the soliton parameters of the system of equations.
\end{itemize}

In the following sections we mainly follow the above method in obtaining the soliton solutions of AKNS($-n$) systems for $n=0, 1,$ and $n=2$
by using Type 1 and Type 2 approaches given in \cite{GurPek2}.

\subsection{One-Soliton Solutions of Local Reduced Equations}

The constraints that one-soliton solutions of the local equations (\ref{localreducedn=0}), (\ref{localreducedn=1}), and (\ref{localreducedn=2})
which are reduced from AKNS($-n$) for $n=0, 1,$ and $n=2$ systems respectively can be found by the local reduction formula $q(x,y,t)=\sigma \bar{p}(x,y,t)$
that is
\begin{equation}\label{relationonelocal}\displaystyle
\frac{e^{k_2x+\tau_2y+\omega_2t+\delta_2}}{1+Ae^{(k_1+k_2)x+(\tau_1+\tau_2)y+(\omega_1+\omega_2)t+\delta_1+\delta_2}}
=\frac{\sigma e^{\bar{k}_1x+\bar{\tau}_1y+\bar{\omega}_1t+\bar{\delta}_1}}{1+\bar{A}e^{(\bar{k}_1+\bar{k}_2)x+(\bar{\tau}_1+\bar{\tau}_2)y+(\bar{\omega}_1+\bar{\omega}_2)t
+\bar{\delta}_1+\bar{\delta}_2}}.
\end{equation}

\noindent If we use the Type 1 approach, we obtain the following constraints:
\begin{equation}
1)\,k_2=\bar{k}_1,\quad  2)\, \omega_2=\bar{\omega}_1,\quad  3)\, e^{\delta_2}=\sigma e^{\bar{\delta}_1},
\end{equation}
so that the equality (\ref{relationonelocal}) is satisfied for each $n=0, 1,$ and $n=2$. Note that under the above constraints, the dispersion
relations give $\tau_2=\bar{\tau}_1$. Hence one-soliton solutions of (\ref{localreducedn=0}), (\ref{localreducedn=1}), and (\ref{localreducedn=2}) are
given by
\begin{equation}\label{localsoln}
p(x,y,t)=\frac{e^{k_1x+\tau_1y+\omega_1t+\delta_1}}{1-\frac{\sigma}{(k_1+\bar{k}_1)^2}
e^{(k_1+\bar{k}_1)x+(\tau_1+\bar{\tau}_1)y+(\omega_1+\bar{\omega}_1)t+\delta_1+\bar{\delta}_1}},
\end{equation}
where
\begin{itemize}
\item[i.] for $n=0$, $a$ and $b$ are pure imaginary numbers and $\tau_1=\frac{1}{b}(\frac{1}{2}k_1\omega_1-ak_1)$,
\item[ii.] for $n=1$, $a$ is a real and $b$ is a pure imaginary number and $\tau_1=\frac{1}{b}(\frac{1}{2}k_1\omega_1-\frac{a}{2}k_1^2)$,
\item[iii.]  for $n=2$, $a$ and $b$ are pure imaginary numbers and $\tau_1=\frac{1}{b}(\frac{1}{2}k_1\omega_1-\frac{a}{4}k_1^3)$.
\end{itemize}

\noindent If $\mathrm{sign}(\sigma)<0$ we can let
\begin{equation}
\sigma=-(k_{1}+\bar{k}_{1})^2\, e^{\mu},
\end{equation}
where $\mu$ is another real constant. Then the above one-soliton solution becomes
\begin{equation}
p(x,y,t)=\frac{e^{\phi}}{2 \cosh \theta}
\end{equation}
where
\begin{eqnarray}
&&\theta= \frac{1}{2}[(k_{1}+\bar{k}_{1}) x+(\tau_{1}+\bar{\tau}_{1}) y+(w_{1}+\bar{w}_{1}) t+\delta_{1}+\bar{\delta}_{1}+\mu],\\
&&\phi = \frac{1}{2}[(k_{1}-\bar{k}_{1}) x+(\tau_{1}-\bar{\tau}_{1}) y+(w_{1}-\bar{w}_{1}) t+\delta_{1}-\bar{\delta}_{1}-\mu],
\end{eqnarray}
Hence one-soliton solutions of the locally reduced equations for $n=0,1,2$ are finite and bounded when $\mathrm{sign}(\sigma)<0$. The norm of $p$ becomes

\begin{equation}
|p(x,y,t)|^2=\frac{e^{-\mu}}{4\cosh^2 \theta}.
\end{equation}

\subsection{One-Soliton Solutions of Nonlocal Reduced Equations}

Firstly let us consider the nonlocal reduction $q(x,y,t)= \sigma p(\epsilon_1x,\epsilon_2y,\epsilon_3t)$. Here the constraints that one-soliton solutions of the nonlocal equations (\ref{nonlocalredn=0a}), (\ref{nonlocalredn=1a}), and (\ref{nonlocalredn=2a})
which are reduced from AKNS($-n$) for $n=0, 1,$ and $n=2$ systems respectively can be found by
\begin{equation}\label{relationonenonlocala}\displaystyle
\frac{e^{k_2x+\tau_2y+\omega_2t+\delta_2}}{1+Ae^{(k_1+k_2)x+(\tau_1+\tau_2)y+(\omega_1+\omega_2)t+\delta_1+\delta_2}}
=\frac{\sigma e^{\epsilon_1k_1x+\epsilon_2\tau_1y+\epsilon_3\omega_1t+\delta_1}}{1+Ae^{\epsilon_1(k_1+k_2)x+\epsilon_2(\tau_1+\tau_2)y+\epsilon_3(\omega_1+\omega_2)t
+\delta_1+\delta_2}},
\end{equation}
where $A=-\frac{1}{(k_1+k_2)^2}$ and $\tau_i$, $i=1, 2$ can be written in terms of $k_i$ and $\omega_i$ due to the dispersion relations of each case $n=0, 1, 2$.

\noindent If we use the Type 1 approach, we obtain
\begin{equation}
1)\, k_2=\epsilon_1 k_1,\quad 2)\, \omega_2=\epsilon_3 \omega_1,\quad 3)\, e^{\delta_2}=\sigma e^{\delta_1}.
\end{equation}
When we use these constraints with the possibilities for $(\epsilon_1,\epsilon_2,\epsilon_3)$ given in Sects. 7.4, 7.5, and 7.6 on the dispersion relations of the cases $n=0, 1, 2$, we get $\tau_2=\epsilon_2 \tau_1$.\\

\noindent For $n=0$ we have $(\epsilon_1,\epsilon_2,\epsilon_3)=(1,1,-1)$ and one-soliton solution of the reduced equation (\ref{nonlocalredn=0a}) is
\begin{equation}\label{nonsoln=0a}
p(x,y,t)=\frac{e^{k_1x+\tau_1y+\omega_1t+\delta_1}}{1-\frac{\sigma}{4k_1^2},
e^{2k_1x+2\tau_1y+2\delta_1}},
\end{equation}
where $\tau_1=\frac{1}{b}(\frac{1}{2}k_1\omega_1-ak_1)$. Assume that all the parameters; $k_1, \omega_1, \delta_1, a$, and $b$ so
$\tau_1$ are real. Let $\sigma=-4 k_{1}^2 e^{2\mu}$ then
\begin{equation}\label{eqn1}
p(x,y,t)=\frac{e^{\phi}}{1+e^{2 \theta+2 \mu}},
\end{equation}
where $\mu$ is a real constant and
\begin{eqnarray}
&&\phi=k_{1} x+\tau_{1} y+ \omega_{1} t+ \delta_{1}, \\
&&\theta=k_{1} x+\tau_{1} y+\delta_{1}.
\end{eqnarray}
Eq. (\ref{eqn1}) can further be simplified as
\begin{equation}\label{eqn2}
p(x, y, t)=\frac{e^{\omega_{1} t-\mu}}{2 \cosh (\theta+\mu)}.
\end{equation}
Hence for the defocusing case, $\mathrm{sign}(\sigma)<0$, one-soliton solution is bounded for all $t \ge 0$ for $\omega_1\leq 0$ and finite for all $(x,y,t)$.\\

\noindent For $n=1$ we have $(\epsilon_1,\epsilon_2,\epsilon_3)=(1,-1,1)$ and one-soliton solution of the reduced equation (\ref{nonlocalredn=1a}) is
\begin{equation}\label{nonsoln=1a}
p(x,y,t)=\frac{e^{k_1x+\tau_1y+\omega_1t+\delta_1}}{1-\frac{\sigma}{4k_1^2}
e^{2k_1x+2\omega_1t+2\delta_1}},
\end{equation}
where $\tau_1=\frac{1}{b}(\frac{1}{2}k_1\omega_1-\frac{a}{2}k_1^2)$.  Assume that all the parameters; $k_1, \omega_1, \delta_1, a$, and $b$ so
$\tau_1$ are real. Let $\sigma=-4 k_{1}^2 e^{2\mu}$ then
\begin{equation}\label{eqn3}
p(x,y,t)=\frac{e^{\phi}}{1+e^{2 \theta+2 \mu}}
\end{equation}
where $\mu$ is a real constant and
\begin{eqnarray}
&&\phi=k_{1} x+\tau_{1} y+ \omega_{1} t+ \delta_{1}, \\
&&\theta=k_{1} x+\omega_{1} t+\delta_{1},
\end{eqnarray}
which can be simplified as
\begin{equation}\label{eqn4}
p(x,y,t)=\frac{e^{\tau_{1} y-\mu}}{2 \cosh (\theta+\mu)}.
\end{equation}
Hence for $\mathrm{sign}(\sigma)<0$, one-soliton solution is finite for all $(x,y,t)$ but not bounded.\\

\noindent For $n=2$ we have $(\epsilon_1,\epsilon_2,\epsilon_3)=(1,1,-1)$ and one-soliton solution of the reduced equation (\ref{nonlocalredn=2a}) is
\begin{equation}\label{eqn5}
p(x,y,t)=\frac{e^{k_1x+\tau_1y+\omega_1t+\delta_1}}{1-\frac{\sigma}{4k_1^2}
e^{2k_1x+2\tau_1y+2\delta_1}},
\end{equation}
where $\tau_1=\frac{1}{b}(\frac{1}{2}k_1\omega_1-\frac{a}{4}k_1^3)$. Hence, similar to $n=0$ case, the solution (\ref{eqn5}) can be simplified
to the form (\ref{eqn2}) with only difference in $\tau_1$. And that solution is bounded for all $t \ge 0$ for $\omega_1\leq 0$ and finite for all $(x,y,t)$ when $\mathrm{sign}(\sigma)<0$.\\

\noindent Note that one of the possibilities in each of the cases for $(\epsilon_1,\epsilon_2,\epsilon_3)$ is $(-1,-1,-1)$. Clearly, because of the definition of the constant $A$, if we use Type 1 approach we obtain trivial solution. Hence we use Type 2 on
\begin{equation}\displaystyle
\frac{e^{\theta_2}}{1+Ae^{\theta_1+\theta_2}}=\sigma\frac{e^{\theta_1^{-}}}{1+Ae^{\theta_1^{-}+\theta_2^{-}}}.
\end{equation}
From the application of the cross multiplication we get
\begin{equation}
e^{\theta_2}+Ae^{2\delta_2}e^{\theta_1^{-}}=ke^{\theta_1^{-}}+Ake^{2\delta_1}e^{\theta_2},
\end{equation}
where
\begin{equation*}
\theta_j=k_jx+\tau_j y+\omega_jt+\delta_j,\quad \theta_1^{-}=-k_jx-\tau_j y-\omega_jt+\delta_j,\quad j=1, 2.
\end{equation*}
Hence we obtain the conditions
\begin{equation}
1)\, A\sigma e^{2\delta_1}=1,\quad 2)\, Ae^{2\delta_2}=\sigma,
\end{equation}
yielding $e^{\delta_1}=\xi_1 i\frac{(k_1+k_2)}{\sqrt{\sigma}}$ and $e^{\delta_2}=\xi_2 i\sqrt{\sigma}(k_1+k_2)$ for $\xi_j=\pm 1$, $j=1, 2$. Therefore
one-soliton solution of the equations (\ref{nonlocalredn=0a}), (\ref{nonlocalredn=1a}), and (\ref{nonlocalredn=2a}) is given by
\begin{equation}\label{eqn6}\displaystyle
p(x,y,t)=\frac{i\xi_1 e^{k_1x+\tau_1y+\omega_1t}(k_1+k_2)}{\sqrt{\sigma}(1+\xi_1\xi_2e^{(k_1+k_2)x+(\tau_1+\tau_2)y+(\omega_1+\omega_2)t})}, \quad \xi_j=\pm 1, j=1, 2,
\end{equation}
with corresponding dispersion relations; (\ref{dispersionn=0ONE}) for $n=0$, (\ref{dispersionn=1ONE}) for $n=1$, and (\ref{dispersionn=2ONE}) for $n=2$.
We can further simplify the solution (\ref{eqn6}) as
\begin{equation}\label{eqn7}\displaystyle
p(x,y,t)=\frac{e^{\phi+\delta_1}}{2\cosh\theta},
\end{equation}
where
\begin{align}
&\phi=\frac{1}{2}[(k_1-k_2)x+(\tau_1-\tau_2)y+(\omega_1-\omega_2)t],\\
&\theta=\frac{1}{2}[(k_1+k_2)x+(\tau_1+\tau_2)y+(\omega_1+\omega_2)t].
\end{align}
The solution (\ref{eqn7}) is finite if $k_1+k_2$, $\tau_1+\tau_2$, and $\omega_1+\omega_2$ are real. In addition to that it is bounded if
$k_1-k_2=0$, $\tau_1-\tau_2=0$, and $\omega_1-\omega_2\leq 0$ for $t\geq 0$. For $n=0$ and $n=2$ cases, these conditions are satisfied
if $k_1, \tau_1, \omega_1$ are real, $k_1=k_2, \tau_1=\tau_2, \omega_2=-\omega_1$, and $\omega_1 \leq 0$ for $t\geq 0$. For $n=1$ case, they
are satisfied if $k_1, \tau_1, \omega_1, a$ are real, $k_1=k_2, \tau_1=\tau_2, \omega_2=2ak_1-\omega_1$, and $\omega_1-ak_1 \leq 0$ for $t\geq 0$.

\bigskip

\noindent The second nonlocal reduction formula is $q(x,y,t)= \sigma \bar{p}(\epsilon_1x,\epsilon_2y,\epsilon_3t)$. The constraints that one-soliton solutions of the nonlocal equations (\ref{nonlocalredn=0b}), (\ref{nonlocalredn=1b}), and (\ref{nonlocalredn=2b})
which are reduced from AKNS($-n$) for $n=0, 1,$ and $n=2$ systems respectively can be found by
\begin{equation}\label{relationonenonlocalb}\displaystyle
\frac{e^{k_2x+\tau_2y+\omega_2t+\delta_2}}{1+Ae^{(k_1+k_2)x+(\tau_1+\tau_2)y+(\omega_1+\omega_2)t+\delta_1+\delta_2}}
=\frac{\sigma e^{\epsilon_1\bar{k}_1x+\epsilon_2\bar{\tau}_1y+\epsilon_3\bar{\omega}_1t+\bar{\delta}_1}}{1+\bar{A}e^{\epsilon_1(\bar{k}_1+\bar{k}_2)x+\epsilon_2(\bar{\tau}_1+\bar{\tau}_2)y
+\epsilon_3(\bar{\omega}_1
+\bar{\omega_2})t
+\bar{\delta}_1+\bar{\delta}_2}},
\end{equation}
where $A=-\frac{1}{(k_1+k_2)^2}$ and $\tau_i$, $i=1, 2$ satisfy the dispersion relations given for each case $n=0, 1, 2$.

\noindent By applying the Type 1 approach, we obtain
\begin{equation}
1)\, k_2=\epsilon_1 \bar{k}_1,\quad 2)\, \omega_2=\epsilon_3 \bar{\omega}_1,\quad 3)\, e^{\delta_2}=\sigma e^{\bar{\delta}_1}.
\end{equation}
Using these constraints besides the conditions (\ref{addcondn=0}), (\ref{addcondn=1}), and (\ref{addcondn=2}) in the dispersion relations of the cases $n=0, 1, 2$ we get $\tau_2=\epsilon_2 \bar{\tau}_1$.\\

\noindent Thus one-soliton solution of the reduced equations (\ref{nonlocalredn=0b}), (\ref{nonlocalredn=1b}), and (\ref{nonlocalredn=2b}) is given by
\begin{equation}\label{nonlocalsolnb}
p(x,y,t)=\frac{e^{k_1x+\tau_1y+\omega_1t+\delta_1}}{1-\frac{\sigma}{(k_1+\epsilon_1\bar{k}_1)^2}
e^{(k_1+\epsilon_1\bar{k}_1)x+(\tau_1+\epsilon_2\bar{\tau}_1)y+(\omega_1+\epsilon_3\bar{\omega}_1)t+\delta_1+\bar{\delta}_1}},
\end{equation}
with the corresponding dispersion relations $\tau_1=\frac{1}{b}(\frac{1}{2}k_1\omega_1-ak_1)$, $\tau_1=\frac{1}{b}(\frac{1}{2}k_1\omega_1-\frac{a}{2}k_1^2)$, and $\tau_1=\frac{1}{b}(\frac{1}{2}k_1\omega_1-\frac{a}{4}k_1^3)$ given respectively. It is clear that there are finite and singular solutions (\ref{nonlocalsolnb}) depending on the parameters of the solutions.\\

\noindent Note that since there are 21 nonlocal reduced equations by the reduction formula $q(x,y,t)=\sigma \bar{p}(\epsilon_1 x,\epsilon_2 y, \epsilon_3 t)$ for $n=0, 1, 2$ let us only consider $y$-reflection that is when $(\epsilon_1,\epsilon_2,\epsilon_3)=(1,-1,1)$ as an example. Let $\sigma=-(k_1+\bar{k}_1)^2e^{\mu}$, $\mu$ is a real constant. Then one-soliton solutions of the nonlocal equations:
\begin{equation}
(n=0),\quad bp_y(x,y,t)=\frac{1}{2}p_{tx}(x,y,t)-ap_x(x,y,t)-\sigma p(x,y,t)D^{-1}(p(x,y,t)\bar{p}(x,-y,t))_t,\hspace{0.6cm}
\end{equation}
where $a$ is a pure imaginary, $b$ is a real number,
\begin{align}
(n=1),\quad bp_y(x,y,t)=&\frac{1}{2}p_{tx}(x,y,t)-\frac{a}{2}p_{xx}(x,y,t)+a\sigma p^2(x,y,t)\bar{p}(x,-y,t)\hspace{2.8cm}
\nonumber\\&-\sigma p(x,y,t)D^{-1}(p(x,y,t)\bar{p}(x,-y,t))_t,
\end{align}
where $a$ and $b$ are real numbers,
\begin{align}
(n=2),\quad bp_y(x,y,t)=&\frac{1}{2}p_{tx}(x,y,t)-\frac{a}{4}p_{xxx}(x,y,t)+\frac{3a}{2}\sigma p(x,y,t)\bar{p}(x,-y,t)p_x(x,y,t)
\hspace{0.8cm}\nonumber\\&-\sigma p(x,y,t)D^{-1}(p(x,y,t)\bar{p}(x,-y,t))_t,
\end{align}
where $a$ is a pure imaginary, $b$ is a real number, become
\begin{equation}\label{eqn8}\displaystyle
p(x,y,t)=\frac{e^{\phi}}{2\cosh(\theta)},
\end{equation}
where
\begin{align}
&\phi=\frac{1}{2}[(k_1-\bar{k}_1)x+(\tau_1+\bar{\tau}_1)y+(\omega_1-\bar{\omega}_1)t+(\delta_1-\bar{\delta}_1-\mu))],\\
&\theta=\frac{1}{2}[(k_1+\bar{k}_1)x+(\tau_1-\bar{\tau}_1)y+(\omega_1+\bar{\omega}_1)t+(\delta_1+\bar{\delta}_1-\mu))].
\end{align}
The solution (\ref{eqn8}) is finite if $\tau_1-\bar{\tau}_1 \in \mathbb{R}$ which happens when $\tau_1 \in \mathbb{R}$. In addition to that it is bounded
if $k_1-\bar{k}_1=0$, $\tau_1+\bar{\tau}_1=2\tau_1=0$, and $\omega_1-\bar{\omega}_1\leq 0$ for $t\geq 0$. This occurs only when $k_1\in \mathbb{R}$ and
$\tau_1=0$.  But taking $\tau_1=0$ reduces the dimension of the solution from $2+1$ to $1+1$.

\section{Conclusion}

In this work we obtained a new negative AKNS hierarchy denoted by AKNS($-n$) for $n=0, 1, 2, \ldots$ in $2+1$ dimensions. We obtained the Hirota bilinear forms of these systems and found one- and two-soliton solutions for $n=0,1,2$. We then found all possible local and nonlocal reductions of these systems. Using the constraint equations among the dynamical variables  for $n=0,1,2$ we found 3 new local and 27 new nonlocal reduced equations in $2+1$ dimensions.
These new nonlocal equations contain two different types of nonlocality. They contain terms with $D^{-1}$ (integro-differential equations) and terms $p(\epsilon_{1} x, \epsilon_{2} y, \epsilon_{3} t)$ (mirror symmetric terms) where $\epsilon_{1}^2=\epsilon_{2}^2=\epsilon_{3}^2=1$.
From the one-soliton solutions of the negative AKNS system of equations we obtained one-soliton solutions of the local and nonlocal reduced equations.  Among all these one-soliton solutions there are solutions which develop
singularities in a finite time and there are also solutions which are finite and bounded depending on the parameters of the solutions.
\bigskip
\bigskip

\vspace{1cm}

\section{Acknowledgment}
  This work is partially supported by the Scientific
and Technological Research Council of Turkey (T\"{U}B\.{I}TAK).

\end{document}